# A two-sided academic landscape: portrait of highly-cited documents in Google Scholar (1950-2013)

Alberto Martín-Martín[*], Enrique Orduna-Malea[**], Juan M. Ayllón[***] and Emilio Delgado López-Cózar[****]

[*] EC3 Research Group. Universidad de Granada. Correo-e: albertomartin@ugr.es
[**] EC3 Research Group. Universitat Politècnica de València. Correo-e: enorma@upv.es
[***] EC3 Research Group. Universidad de Granada. Correo-e: jayllon@ugr.es
[****] EC3 Research Group. Universidad de Granada. Correo-e: edelgado@ugr.es

**Abstract:** The main objective of this paper is to identify the set of highly-cited documents in Google Scholar and to define their core characteristics (document types, language, free availability, source providers, and number of versions), under the hypothesis that the wide coverage of this search engine may provide a different portrait about this document set respect to that offered by the traditional bibliographic databases. To do this, a query per year was carried out from 1950 to 2013 identifying the top 1,000 documents retrieved from Google Scholar and obtaining a final sample of 64,000 documents, of which 40% provided a free full-text link. The results obtained show that the average highly-cited document is a journal article or a book (62% of the top 1% most cited documents of the sample), written in English (92.5% of all documents) and available online in PDF format (86.0% of all documents). Yet, the existence of errors especially when detecting duplicates and linking cites properly must be pointed out. The fact of managing with highly cited papers, however, minimizes the effects of these limitations. Given the high presence of books, and to a lesser extend of other document types (such as proceedings or reports), the research concludes that Google Scholar data offer an original and different vision of the most influential academic documents (measured from the perspective of their citation count), a set composed not only by strictly scientific material (journal articles) but academic in its broad sense.

## Un panorama académico de dos caras: retrato de los documentos altamente citados en Google Scholar (1950-2013)

**Resumen:** El principal objetivo de este trabajo es identificar el conjunto de documentos altamente citados en Google Scholar y definir sus características nucleares (tipología documental, idioma, disponibilidad en abierto, fuentes y número de versiones), bajo la hipótesis de que la amplia cobertura del buscador podría proporcionar un retrato diferente de este conjunto documental a la ofrecida por las bases de datos tradicionales. Para ello, se ha realizado una consulta por año (desde 1950 hasta 2013) identificando los 1000 documentos más citados y obteniendo una muestra final de 64.000 registros (el 40% de los cuales proporcionaban un enlace al texto completo). Los resultados muestran que el documento altamente citado "promedio" es un artículo de revista o libro (éstos constituyen el 62% del top 1% de los documentos más citados de la muestra), escrito en inglés (92.5%) y disponible online en PDF (86% de la muestra). Aun así, se debe indicar la existencia de errores especialmente en la detección de documentos duplicados y en la correcta vinculación de citas. En todo caso, la muestra manejada (documentos altamente citados) minimiza los efectos de dichas limitaciones. Dada la alta presencia de libros (manuales) y, en menor medida, de otras tipologías documentales (como congresos o informes), se concluye que Google Scholar ofrece una visión original y diferente del conjunto de documentos académicos más influyentes (medidos desde la perspectiva de la contabilización de citas), conformado no sólo por material estrictamente científico (artículos en revistas), sino académico en sentido amplio.

**Palabras clave:** Google Scholar, motores de búsqueda académicos, documentos altamente citados, libros académicos, acceso abierto.

# 1. INTRODUCTION

The idea of identifying the most influential scientific documents using the number of times they are cited in the scientific literature was introduced, like many other bibliometric procedures, by Garfield (1977). The candidates for "Citation classics" were selected from a group of the 500 most cited papers during the years 1961-1975 (http://garfield.library.upenn.edu/classics.html). From 2001, the highly cited papers were integrated in a new product: The Essential Science Indicators. Nevertheless, no other bibliographic database has released alternatives to this product.

However, we do have an extensive scientific literature on the matter of highly-cited documents in different journals, subject areas, institutions or countries (Oppenheim & Renn, 1978; Narin et al., 1983; Plomp, 1990; Glänzel & Czerwon, 1992; Glänzel & Schubert, 1992; Glänzel et al., 1995; Tijssen et al., 2002; Aksnes, 2003; Aksnes & Sivertsen, 2004; Kresge et al., 2005; Levitt & Thelwall, 2009; Smith, 2009; Persson, 2010).

Recently, the interest in these lists has returned with the development of rankings based on the concept of excellence through the calculation of percentiles, first suggested by Garfield (1979), after used by Narin (1987) and Van Raan and Hartmanm (1987), and specified in a systematic proposal to measure excellence by Maltrás (2003), and recently popularized by other authors (Bornmann, 2010; Bornmann & Mutz, 2011; Bornmann et al., 2011).

To celebrate the fiftieth anniversary of the Science Citation Index, the journal *Nature* asked *Thomson Reuters* for the list of the top 100 most highly-cited papers of all time (Van Noorden et al., 2014). In this list, the classic "Protein measurement with the folin phenol reagent", by Lowry et al. (1951), achieves the first position, a place it has historically occupied (Garfield, 2005; Kresge et al., 2005). Although the authors explore the most-cited research of all time using data from the Web of Science Core Collection (WoScc), they also provide an alternative ranking using data from Google Scholar (GS). This alternative ranking is only available in the online version of that article as supplementary material.[1]

The appearance of Google Scholar at the end of 2004 signalled a revolution in the way scientific publications were searched, retrieved and accessed (Jacsó, 2005), becoming not only a search engine for academic documents, but also for the citations these documents receive (Ortega, 2014).

Google Scholar's crawlers systematically parse and analyse the entire academic web, not making distinctions based on subject areas, languages, or countries. Although Google Scholar also achieves agreements with private commercial publishers which may somewhat compromise its neutral coverage, the use of crawlers undoubtedly enables the calculation of impact metrics for a broader collection of documents, not only articles published in elite journals that are included in expensive citation indexes. Disciplines inside the Social Sciences and the Humanities, which use other channels of scientific communication apart from journal articles (such as doctoral theses, books, book chapters, working papers, and conference proceedings) could benefit from using this much broader source of scientific publication data (Meho & Yang, 2007; Harzing & Van der Wal, 2008; Bar-Ilan, 2010; Kousha et al., 2011; Kousha & Thelwall, 2008).

Its wide coverage and evolution (Aguillo, 2012; Khabsa & Giles, 2014; Ortega, 2014; Winter et al., 2014; Orduna-Malea et al., 2015) as well as its empirically tested capacity to obtain unique citations (citations that can only be found in Google Scholar) (Yang & Meho, 2006; Meho & Yang, 2007; Kousha & Thelwall, 2008; Bar-Ilan, 2010; Kousha et al., 2011; Harzing, 2013; Harzing, 2014; Orduna-Malea & Delgado López-Cózar, 2014), make of Google Scholar an exceptional source to collect highly-cited documents.

One issue that should be taken into account when using bibliographic and bibliometric data provided by Google Scholar is that the data may present errors. These errors have been already studied and classified (Jacsó, 2005; 2006; Bar-Ilan 2010; Jacsó 2008a; 2008b; 2012). Although the quality of the data has improved significantly over the years (Google Scholar is now over 11 years old), some of these errors still persist, especially those related to the detection of duplicate documents, and the correct allocation of citations (Martín-Martín et al., 2015; Orduna-Malea et al., 2015). Thus, Google Scholar data usually requires some cleaning before it is suitable for analysis. Failing to observe this measure might lead to unreliable results. This is the case of Nature's ranking of highly cited documents according to Google Scholar (Van Noorden et al., 2014), which presents various irregularities (Martín-Martín et al., 2015).

In spite of these shortcomings, Google Scholar is capable not only of identifying the most-cited papers, but also of providing a view of a broader academic landscape (including books, heavily cited in certain fields, and traditionally discriminated against).

It is important to note that Nature's ranking was drawn from the data that the Google Scholar's team provided directly to the authors. It would be necessary therefore to ascertain whether such listings could be obtained by an independent user through the use of specific search queries. This task has been carried out successfully (see supplementary material), demonstrating the soundness of Google Scholar for retrieving highly-cited documents, and providing an opportunity for the execution of studies describing the key bibliographic aspects of these highly-cited items. The unique coverage policy of Google Scholar (virtually no language, country, subject area, or document type restrictions) may provide interesting insights to the bibliometric community for understanding the characteristics of highly-cited documents.

Although some of the bibliographic properties of the documents indexed in Google Scholar (such as its sources or document types) have been previously treated in the existing literature, these works have never focused on samples made up entirely of highly-cited documents. Aguillo (2012) and Ortega (2014) performed two separate general analyses of the search engine (without considering the number of citations received by documents), while Jamali and Nabavi (2015) studied a sample of 8310 documents in different disciplinary fields (the 277 subcategories offered by Scopus), and limited to the period 2004-2014. In fact, the use of keyword queries prevented the authors from isolating highly-cited papers, since those queries were affected by Google Scholar's academic search engine optimization practices (Beel et al., 2010). This issue is circumvented in this work by means of using keyword-free year queries.

Therefore, the main objective of this paper is to identify the set of highly-cited documents in Google Scholar and define their core characteristics, in order to give an answer to the following research questions:
   - Which are the most cited documents in Google Scholar?

- Which is the most frequent document type for these highly-cited documents?
- In what languages are the most cited documents written?
- How many highly-cited documents are freely accessible?
- What are the most common file formats to store these highly cited documents?
- Which are the main providers of these highly-cited full text documents?
- How many versions has Google Scholar found of these highly-cited documents?

## 2. METHOD

In order to isolate a sample of highly-cited documents, we performed a series of keyword-free year queries (only the year field in Google Scholar's advanced search was used). By doing this, the results of the queries weren't limited to a specific topic.

A longitudinal analysis was carried out by performing 64 keyword-free year queries from 1950 to 2013 (one query per year). All the records displayed (a maximum of 1,000 per query) were extracted, obtaining a final set of 64,000 records.

This process was carried out twice (on the 28$^{th}$ of May, and on the 2$^{nd}$ of June, 2014). In the first case, it was performed from a computer with access to the Web of Science, in order to obtain WoS data embedded in Google Scholar (http://wokinfo.com/googlescholar). In the second case, the data extraction was made from a computer with a normal Internet connection, because we wanted to collect data about free full-text links that couldn't have been unadulterated by our university's subscriptions. This process doubled as a reliability check, because we confirmed that the two datasets contained the same records. After this, the HTML source code for each of the search engine result pages of every query was parsed using a self-elaborated web scraper, extracting all the bibliographic information available for every record (Fig. 1).

**Figure 1. Fields extracted from each Google Scholar record in the search engine results page**

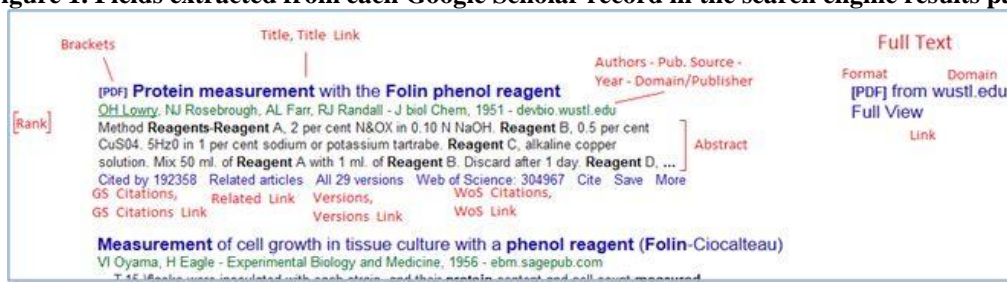

The main fields extracted were the following: author name(s), publication source, year of publication, GS citations, and number of versions.

The full-text fields are available only when Google Scholar finds at least one freely accessible version among all the versions identified of a same document. In the cases where more than one free version is found, Google Scholar selects one of them and displays it right next to the bibliographic information of the primary version of the document. This study analyses only those selected full-text links, not all the full-text links that may be found when clicking the "View all X versions" link of a Google Scholar result. For each document with full-text data, the following fields were extracted: domain (the web domain where GS has found a free full-text version of the document), link (hyperlink to the full-text of the document), and format (file type of the full-text version of the document).

In addition to these fields, information about the document type and the language of the document, which are not systematically provided by Google Scholar, were assigned to each record as well.

Regarding the document types, some records display a text in square brackets before the title of the document (for example "[BOOK]"). Regrettably, this text is not always offered and in some cases the information does not refer to document types but to file types (for example "[PDF]" or "[HTML]") or it is used to mark some special records, such as "[CITATION]", references to a document that have been found cited in the reference list of a document indexed in Google Scholar, but are not linked to any web source. Since some "citation records" are actually versions not properly linked to main records (and may contain additional cites), these have been captured and processed.

Given the difficulty of ascertaining the typologies of the documents indexed in Google Scholar, we devised three different strategies that, combined, allowed us to some extent to define the typology of the documents in the data set:
  a) All documents where the field brackets = "[BOOK]" were considered as books (codified as "B").
  b) For documents that were also indexed in WoS, Google Scholar data was merged with WoS data to obtain the document types. The correspondence is as follows:
      - Journal ("J"): "Article", "Letter", "Note", "Reviews".
      - Book ("B"): "Book", "Book Chapter".
      - Conference proceedings ("C"): "Proceedings Papers".
      - Others ("O"): "Book Review", "Correction", "Correction, Addition", "Database Review", "Discussion", "Editorial Material", "Excerpt", "Meeting Abstract", "News Item", "Poetry", "Reprint", "Software Review".
  c) Lastly, we analyzed the publication source (where possible), searching for keywords (in different languages) that could indicate the type of the source publication, searching the following terms:
      - Journal ("J"): "Revista", "Anuario", "Cuadernos", "Journal", "Revue", "Bulletin", "Annuaire", "Anales", "Cahiers", "Proceedings".
      - Conference Proceedings ("C"): "Proceedings", "Congreso", "Jornada", "Seminar", "Simposio", "Congrès", "Conference", "symposi", "meeting".

Since the word "Proceedings" is used both for journals (i.e. "Proceedings of the National Academy of Sciences") and for conference proceedings (i.e., "Proceedings of the 4th Conference…"), records containing this word in the publication source field were all considered initially as conference proceedings, but a manual check was carried out to reassign those that were really journal articles.

With respect to the language of the documents (GS doesn't provide this information either), we manually checked the language in which the title and abstract of the document were written as well as WoS data (when available), as a basis to fill the language field.

Lastly, all the data was saved to a spreadsheet so it could be statistically analyzed. Pearson and Spearman correlations (α=0.01) were calculated with the XLstat statistical suite in order to find the connection between versions and citations.

## 3. RESULTS

**The most cited documents in Google Scholar**

The top 25 most cited documents in GS (1950-2013) are listed in Table I. In the case of books, the year of publication is the year of publication of the first edition. The top 1% most cited documents in our sample (640 documents) are provided in the supplementary material.[1]

**Table I. Top 25 most-cited documents in Google Scholar (1950-2013)**

| R | DOCUMENT (Author, Title, Publisher) | YEAR (1ST ED.) | CITATIONS | TYPE |
|---|---|---|---|---|
| 1 | LOWRY, O.H. et al. Protein measurement with the Folin phenol reagent. *The Journal of biological chemistry*. | 1951 | 253,671 | J |
| 2 | LAEMMLI, U.K. Cleavage of structural proteins during the assembly of the head of bacteriophage T4. *Nature*. | 1970 | 221,680 | J |
| 3 | BRADFORD, M.M. A rapid and sensitive method for the quantitation of microgram quantities of protein using the principle of protein Dye binding. *Analytical Biochemistry*. | 1976 | 185,749 | J |
| 4 | SAMBROOK, J., FRITSCH, E. F., & MANIATIS, T. Molecular cloning: a laboratory manual. New York, Cold Spring Harbor Laboratory Press. | 1982 | 171,004 | B |
| 5 | AMERICAN PSYCHIATRIC ASSOCIATION. Diagnostic and statistical manual: mental disorders. Washington, American Psychiatric Assn. | 1952 | 129,473 | B |
| 6 | PRESS, W. H. Numerical recipes: the art of scientific computing. Cambridge: Cambridge University Press. | 1986 | 108,956 | B |
| 7 | YIN, R. K. Case study research: design and methods. Beverly Hills (CA): Sage Publications. | 1984 | 82,538 | B |
| 8 | ABRAMOWITZ, M., & STEGUN, I. A. Handbook of mathematical functions: with formulas, graphs, and mathematical tables. Washington, Government printing office. | 1964 | 80,482 | B |
| 9 | KUHN, T. S. The structure of scientific revolutions. Chicago, University of Chicago Press. | 1962 | 70,662 | B |
| 10 | ZAR, J. H. Biostatistical analysis. Englewood Cliffs: Prentice Hall international. | 1974 | 68,267 | B |
| 11 | SHANNON, C.E. A mathematical theory of communication. *The Bell System Technical Journal*. | *1948 | 66,851 | J |
| 12 | CHOMCZYNSKI & SACCHI, N. Single-step method of RNA isolation by acid guanidinium thiocyanate-phenol-chloroform extraction. *Analytical Biochemistry* | 1987 | 63,871 | J |
| 13 | SANGER F, NICKLEN S, & COULSON AR. DNA sequencing with chain-terminating inhibitors. *Proceedings of the National Academy of Sciences of the United States of America*. | 1977 | 63,767 | J |
| 14 | COHEN, J. Statistical power analysis for the behavioral sciences. New York: Academic Press. | 1969 | 63,766 | B |
| 15 | GLASER, B. G., & STRAUSS, A. L. The discovery of grounded theory: strategies for qualitative research. New York: Aldine de Gruyter. | 1967 | 61,158 | B |
| 16 | NUNNALLY, J. C. Psychometric Theory. New York: McGraw-Hill. | 1967 | 60,725 | B |
| 17 | GOLDBERG, D. E. Genetic algorithms in search, optimization, and machine learning. Reading, Mass: Addison-Wesley. | 1989 | 59,764 | B |
| 18 | ROGERS, E. M. Diffusion of Innovations. Pxiii. 367. Free Press of Glencoe, New York: Macmillan. | 1962 | 55,738 | B |
| 19 | BECKE, A.D. Density Functional Thermochemistry III The Role of Exact Exchange. *J. Chem. Phys*. | 1993 | 54,642 | J |
| 20 | LEE, C., YANG, W. & PARR, R.G. Development of the Colle-Salvetti correlation-energy formula into a functional of the electron | 1988 | 52,316 | J |
| | density. *Physical Review B*. | | | |
| 21 | MURASHIGE, T. & SKOOG, F. A revised medium for rapid growth and bio assays with tobacco tissue cultures. *Physiologia Plantarum*. | 1962 | 52,011 | J |
| 22 | ANDERSON, B. R. O. Imagined communities: reflections on the origin and spread of nationalism. London: Verso. | 1983 | 51,177 | B |
| 23 | FOLSTEIN, M.F., FOLSTEIN, S.E. & MCHUGH, R. Mini-mental state. *Journal of Psychiatric Research*. | 1975 | 51,150 | J |
| 24 | TOWBIN, H., STAEHELIN, T. & GORDON, J. Electrophoretic transfer of proteins from polyacrylamide gels to nitrocellulose sheets: procedure and some applications. *Proceedings of the National Academy of Sciences of the United States of America*. | 1979 | 50,608 | J |
| 25 | PAXINOS, G., & WATSON, C. The rat brain in stereotaxic coordinates. Sydney [etc.]: Academic Press. | 1982 | 50,471 | B |

J: Article journal; B: Book;
\* Contribution published outside the studied timeframe; fully commented on in the discussion.

The most cited document according to GS is the aforementioned article by Lowry et al, with 253,671 citations (as of May 2014), followed by Laemmly's article, with 221,680 citations.

Although the ranking is dominated by studies from the natural sciences (especially the life sciences), it also contains many works from the social sciences (especially from economics, psychology and sociology), and also from the humanities (philosophy and history). For instance, within the top 20 documents we can find "The structure of scientific revolutions (9[th] position; 70,662 citations) and "Diffusion of innovations" (18[th]; 55,738 citations).

Many of the works in this ranking are methodological in nature: they describe the steps of a certain procedure or how to handle basic tools to process and analyse data. This is exemplified by the presence of statistical manuals ("Handbook of Mathematical Functions with Formulas"), laboratory manuals ("Molecular cloning: a laboratory manual"), manuals of research methodology ("Case study research: design and methods"), and works that have become a de facto standard in professional practice ("Diagnostic and statistical manual of mental disorders").

In fact, books are the most common category among the top 1% most cited documents, constituting the 62% (395) of this subsample, followed by journal articles with 36.01% (231). Moreover, the citation average of books (2,700) is higher than that for journal articles (1,700).

**Document types**

The document type has been identified in 71% (45,440) of the documents sampled, whereas the typology of the other 29% (18,590) remained unknown (our automatic strategies weren't able to determine it, and manual checking would have been too costly). The distribution of document typologies is displayed in Figure 2, where we find a clear predominance of journal articles (including reviews, letters and notes as well) which represent 51% of the total 64,000 documents (72.3% of the documents with a defined document type). Book and book chapters together also make up a big part of the sample (18%; 11,240 items) while the presence of conference proceedings and other typologies (meeting abstracts, corrections, editorial material, etc.) is merely testimonial (1% each).

**Figure 2. Document types of the highly cited documents in Google Scholar**

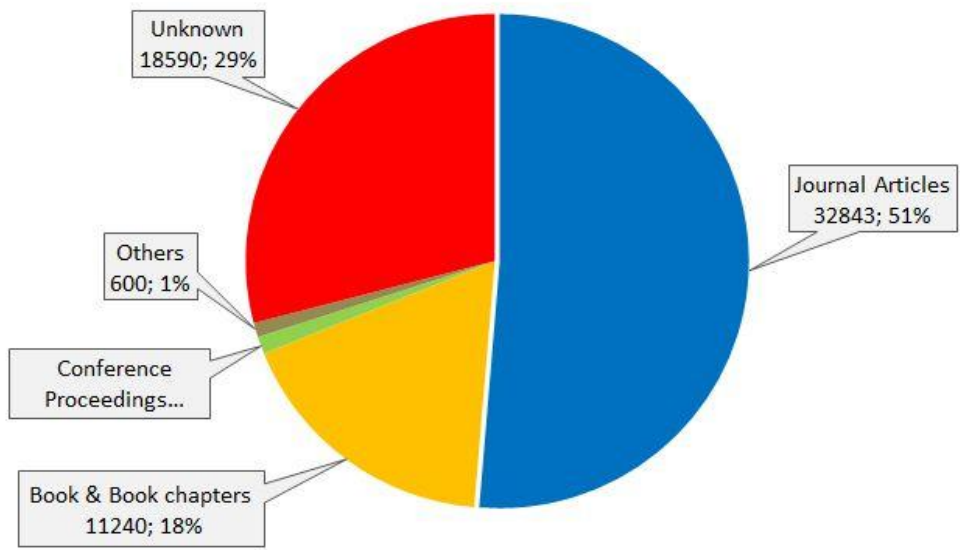

Figure 3 represents this distribution in a longitudinal perspective, where we can observe the following three phenomena:
- Conference proceedings and "Others" categories play an insignificant role along the years, although they achieve greater presence during the last decade.
- A steady decrease over time in the number of documents with an unknown typology (from 35.4% in 1950 to 12.9% in 2013).
- A constant increase in the number of books, which become the most frequent document type in the last five years (2009-2013), monopolizing the sample. As an example, within the 1,000 results corresponding for the year 2013, we only find 27 journal articles but 842 books. The reason for this overrepresentation of the book format in the most recent years is explained in the discussion section of this article.

**Figure 3. Document types of the highly cited documents in Google Scholar, broken down by years (1950-2013)**

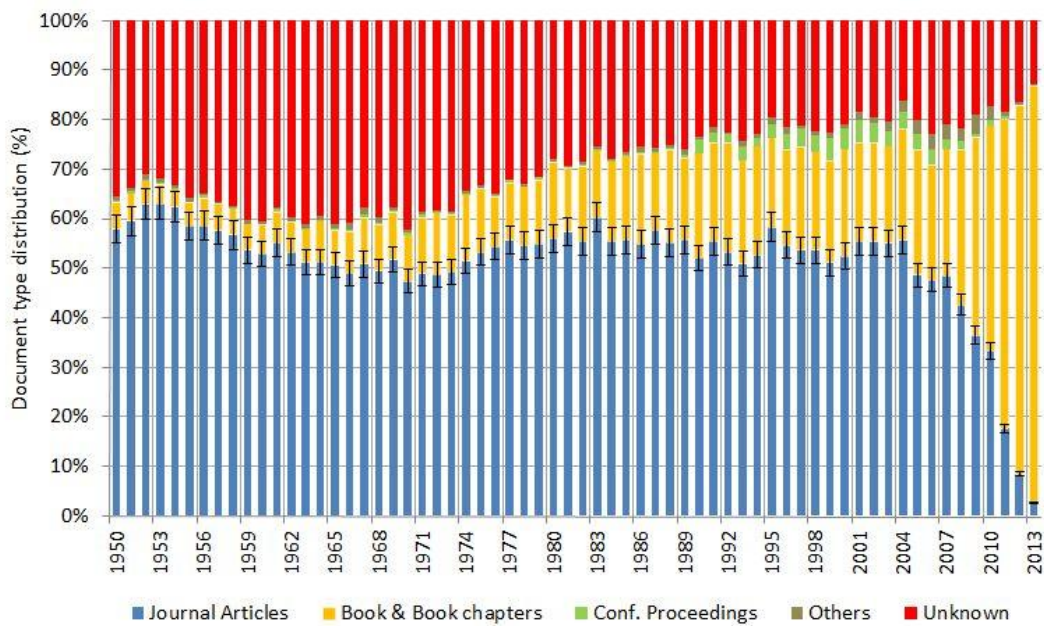

**Language of documents**

In Figure 4 we find the document distribution according to language. As we can see, English dominates over the rest of the languages as the most widely used language for scientific communication in Google Scholar, accounting for 92.5% of all the documents. The second and third places are occupied by Spanish and Portuguese respectively, but neither of them reaches even 2% of the total.

**Figure 4. Distribution of languages used in the highly-cited documents in GS**

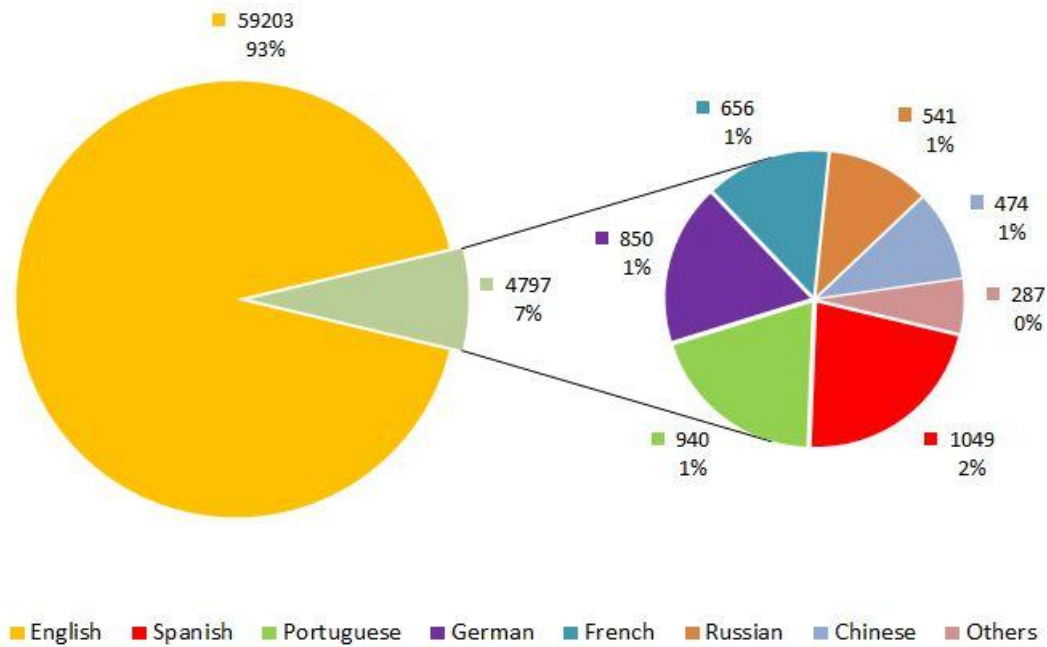

In Figure 5 we can observe the longitudinal evolution of the language usage distribution, which is much more stable through the years than the ones previously found for the document types. The English language predominates during the whole period ($\tilde{x}$= 92.5%; σ= 1.6%), with an oscillation of less than 10% between its maximum and minimum value (87% in 2013, and 95% in 1991). Data also shows a slightly decrease in English percentage in the last three years (from 92% in 2010 to 87.1% in 2013), though more data is required to determine if this change is just circumstantial or a new trend.

The "Others" category (which represents 7% of the documents) includes the following languages: Italian, Swedish, Indonesian, Finnish, Danish, Bulgarian, Polish, Norwegian, Turkish, Latin, Slovenian, Serbian, Dutch, Macedonian, Malayan, Japanese, Czech, Estonian, Slovak, Mongolian, Catalan, Croatian, Lithuanian, and Ukrainian.

**Figure 5. Distribution of languages in the highly cited documents in GS by years of publication (1950-2013)**

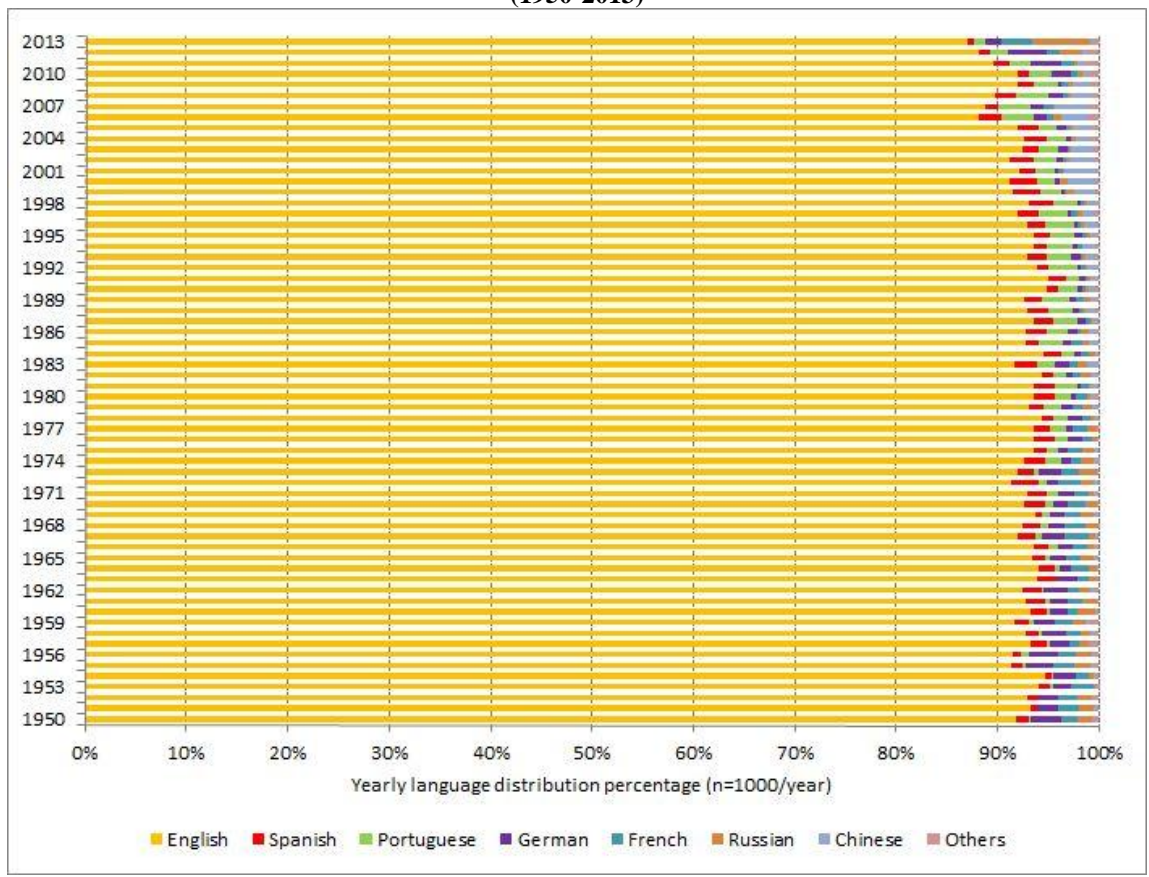

**Availability of Full text documents**

A free full-text link is provided for 40% (25,849) of all the highly-cited documents retrieved (Figure 6; top). We can also observe a positive trend through the analyzed period (from 25.93% of documents with free full-text links in the period 1950-1959, to 66.84% in 2000-2009), although this trend is interrupted in the last four years (41.5% from 2010 to 2013), where the high percentage of books in these years are affecting the results (Figure 6; bottom). The journals' and publishers' embargo policies may have slight influence as well, especially for the experimental sciences.

**Figure 6. Percentage of freely accessible highly cited documents in Google Scholar (1950-2013). Global results (top); broken down by decades (bottom).**

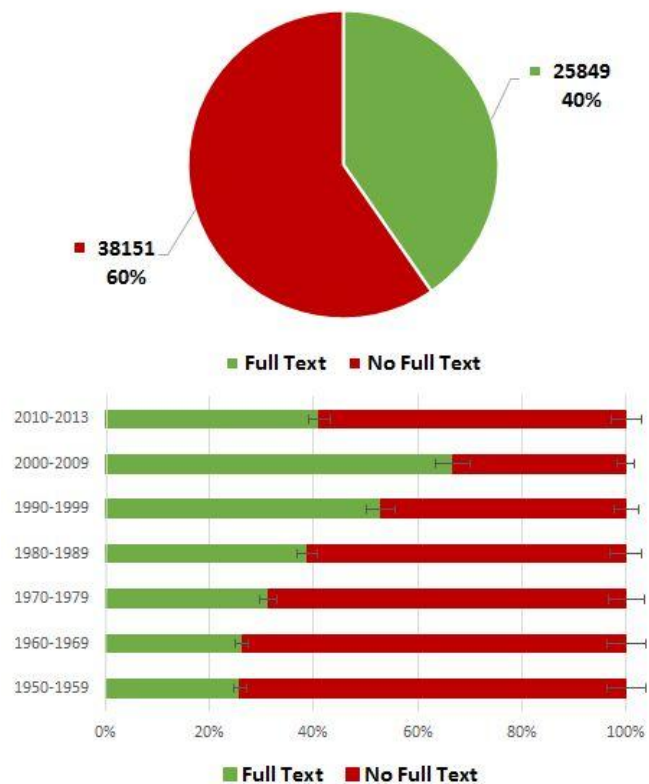

## File types

Full-text links point to documents in a variety of formats (Figure 7). The most common one is the pdf format (86.0% of all full text documents), followed by the html format (12.1%). The remaining identified file formats (doc, ps, txt, rtf, xls, ppt) together only represent 1.9% of the freely available documents.

**Figure 7. File Formats of the freely accessible highly cited documents in Google Scholar (1950-2013)**

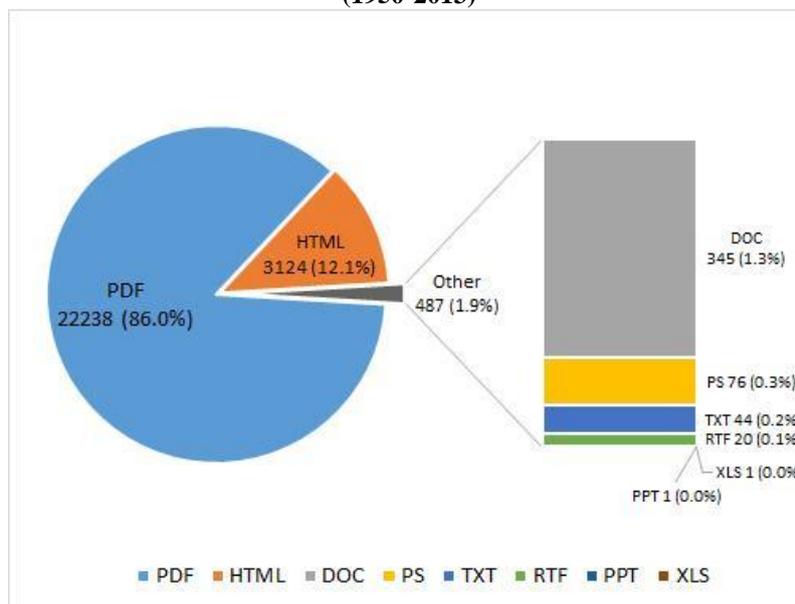

Figure 8 shows the same data broken down by years (1950-2013). We can see that the predominance of the pdf format is patent throughout the entire range of years. However, it is also noteworthy that the html format has started gaining more presence for documents published in the last 25 years, with a peak of almost 20% of the share in 2010.

**Figure 8. File Format distribution for the freely accessible highly cited documents in Google Scholar broken down by years (1950-2013)**

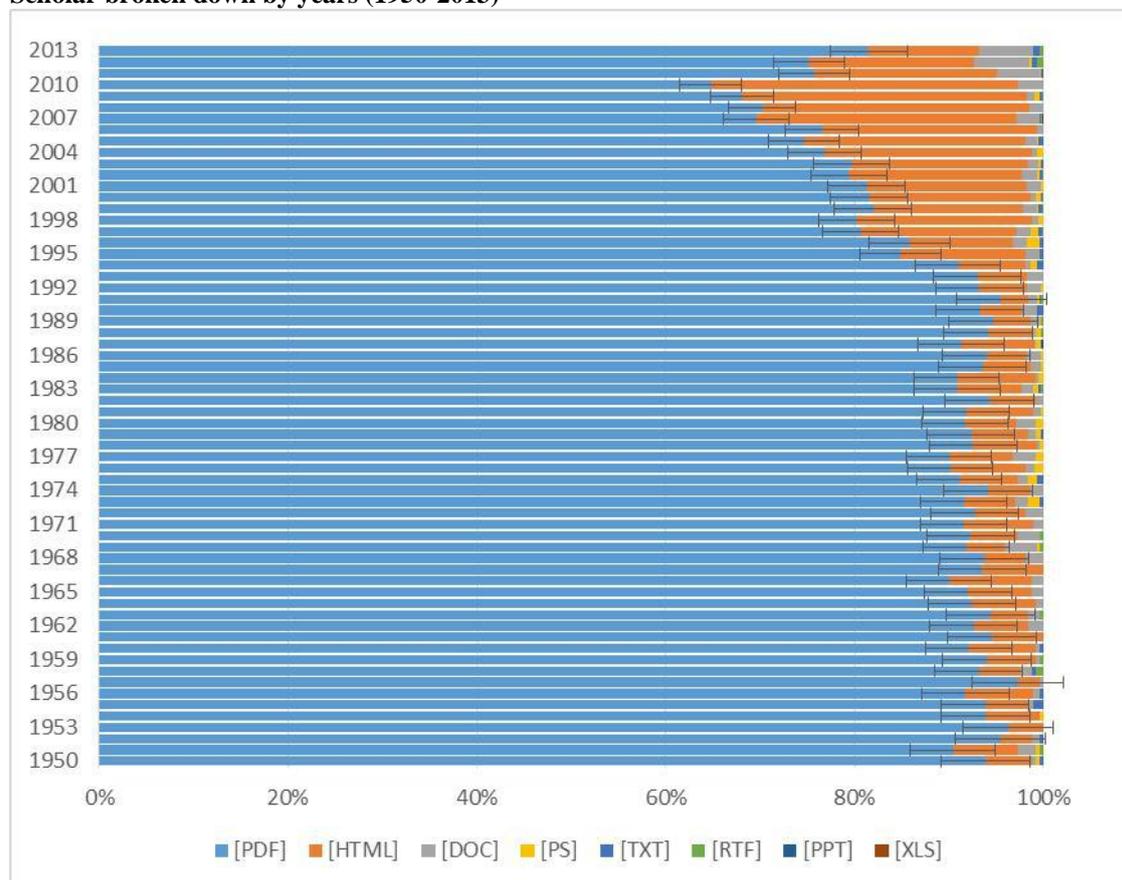

**Full-text source providers**

A total of 5,715 different providers of free full-text links to highly cited documents have been found in the sample. However, a group of 35 providers (18 universities; 5 scientific societies; 4 publishers; 2 companies; 2 public administrations; 1 journal; 1 digital library; 1 repository; 1 academic social network) account for more than a third of all the links (37%).

Table II shows the main providers. The National Institutes of Health (NIH) hold the first position (1,405 documents), mainly due to the Pubmed central repository, hosted within the NIH website (http://www.ncbi.nlm.nih.gov/pubmed). The second position is occupied by ResearchGate (815), followed by Harvard University (495).

**Table II. Top full text source providers in Google Scholar (1950-2013)**

| Provider | Nº | Type of entity |
|---|---|---|
| nih.gov | **1,405** | Public administration |
| researchgate.net | **815** | Academic Social network |
| harvard.edu | **495** | University |
| pnas.org | **478** | Scientific society |
| oxfordjournals.org | **466** | Publisher |
| psu.edu | **424** | University |
| arxiv.org | **423** | Repository |
| jbc.org | **414** | Journal |
| sciencedirect.com | **394** | Publisher |
| wiley.com | **324** | Publisher |
| jstor.org | **322** | Digital library |
| rupress.org | **304** | University |
| royalsocietypublishing.org | **266** | Scientific society |
| ahajournals.org | **218** | Scientific society |
| dtic.mil | **208** | Public administration |
| stanford.edu | **203** | University |
| google.com | **188** | Company |
| mit.edu | **180** | University |
| tu-darmstadt.de | **177** | University |
| nature.com | **161** | Publisher |
| yale.edu | **141** | University |
| caltech.edu | **140** | University |
| physoc.org | **140** | Scientific society |
| cmu.edu | **122** | University |
| umich.edu | **120** | University |
| duke.edu | **118** | University |
| princeton.edu | **116** | University |
| wisc.edu | **113** | University |
| ucsd.edu | **112** | University |
| asm.org | **112** | Scientific society |
| berkeley.edu | **107** | University |
| upenn.edu | **104** | University |
| washington.edu | **103** | University |
| columbia.edu | **102** | University |
| yimg.com | **101** | Company |
| **TOTAL** | **9,616** | |

If we analyse the top-level domains of the 25,849 links to full text available documents (Table III), the most frequent are academic institutions (.edu; 23.74%) and organizations (.org; 21.39%). Moreover, the number of links provided by academic institutions is likely to be higher since there are many universities that use national top-level domains instead of .edu (mostly reserved for North American academic institutions). For example, Technische Universität Darmstadt (tu-darmstadt.de) provides 177 links. At a national scale, some countries use a ""ac.xx" pattern domain, such as United Kingdom (ac.uk), which provides 333 links. The most important geographic domain is Germany (.de) with only 2.62% (678) of the highly-cited documents.

**Table III. Top-level domains providing full text links in Google Scholar (1950-2013)**

| Domain | N | % |
|---|---|---|
| .edu | 6,136 | 23.74 |
| .org | 5,528 | 21.39 |
| .com | 3,466 | 13.41 |
| .gov | 1,712 | 6.62 |
| .net | 1,345 | 5.20 |
| .de | 678 | 2.62 |
| .cn | 489 | 1.89 |
| .uk | 485 | 1.88 |
| .ca | 404 | 1.56 |
| .ru | 374 | 1.45 |
| .fr | 357 | 1.38 |
| .br | 343 | 1.33 |
| .it | 275 | 1.06 |
| .ch | 214 | 0.83 |
| .mil | 210 | 0.81 |
| .nl | 186 | 0.72 |
| .es | 145 | 0.56 |
| .tw | 136 | 0.53 |
| .au | 131 | 0.51 |
| .in | 118 | 0.46 |
| Others | 3,117 | 12.06 |
| **TOTAL** | **25,849** | **100%** |

**Versions**

83.17% (53,229) of the documents analyzed have more than one version (Table IV). The distribution of the number of versions is asymmetric, led by documents with 1 version (16.83; 10,771 documents) and followed by documents with 3 versions (6,903; 10.79%) and 4 versions (6,814; 10.65%). The existence of documents with a massive number of versions is also worth noting. For 281 documents, Google Scholar has found more than 100 versions, and more than 500 versions for 14 of those documents. The document with the highest number of versions in our sample has 899 versions.

**Table IV. Distribution of documents according to their number of versions**

| Nº of versions | Nº of documents | % | Accumulated (docs) | Accumulated (%) |
|---|---|---|---|---|
| 1 | 10,771 | 16.83 | 10,771 | 16.83 |
| 2 | 6,075 | 9.49 | 16,846 | 26.32 |
| 3 | 6,903 | 10.79 | 23,749 | 37.11 |
| 4 | 6,814 | 10.65 | 30,563 | 47.75 |
| 5 | 5,539 | 8.65 | 36,102 | 56.41 |
| 6 | 4,781 | 7.47 | 40,883 | 63.88 |
| 7 | 3,746 | 5.85 | 44,629 | 69.73 |
| 8 | 2,940 | 4.59 | 47,569 | 74.33 |
| 9 | 2,429 | 3.80 | 49,998 | 78.12 |
| 10 | 1,929 | 3.01 | 51,927 | 81.14 |
| 11-15 | 5,243 | 8.19 | 57,170 | 89.33 |
| 16-25 | 3,585 | 5.60 | 60,755 | 94.93 |
| 26-50 | 2,202 | 3.44 | 62,957 | 98.37 |
| 51-100 | 762 | 1.19 | 63,719 | 99.56 |
| 101-200 | 202 | 0.32 | 63,921 | 99.88 |
| 201-300 | 40 | 0.06 | 63,961 | 99.94 |
| 301-400 | 16 | 0.03 | 63,977 | 99.96 |
| 401-500 | 9 | 0.01 | 63,986 | 99.98 |
| > 501 | 14 | 0.02 | 64,000 | 100 |

Pearson's correlation coefficient between the number of citations of a document in Google Scholar and its number of versions is low (r = 0.2; α= 0.01). However, the Spearman correlation shows a better correlation (r= 0.48; α= 0.01). This may be an effect of the highly skewed distribution of citations. For example, the average of citations for documents with at least 100 versions is high (5,878.13), although the Pearson's correlation of these highly-versioned documents with the corresponding number of citations is even lower (r= 0.13).

## 4. DISCUSSION

An in-depth discussion of this radiography of highly-cited documents in Google Scholar is necessary, due to the limitations of the database. We will first consider the key parameters that may have influenced the ranking presented in Table I (essentially the dynamic of citations received, and the number of versions). Next, we'll warn about some flaws that affect the composition of the sample (related to the publication date and the language of the documents). Lastly, we will comment on some specific properties of the documents in our sample (document types, full text, file formats, and providers).

**Key parameters**

*The fluctuation of citations*

In this section we set aside the issues regarding the quality and the source of the citations received by the 64,000 documents analyzed, and the well-known errors related to the inaccurate attribution of citations (which is not so important when we are studying highly-cited documents). Instead, we will focus on an issue which might significantly distort the results of this kind of studies: the fluctuation of citations in Google Scholar.

Unlike in other bibliographic databases (such as Scopus or Web of Science core collection), Google Scholar reflects the number of citations considering the documents that are available on the Web at the time the search is made. Google Scholar's team warns that the database "reflects the state of the web as it is currently visible to our search robots and to the majority of users" (http://scholar.google.com/intl/en/scholar/help.html#corrections). This means that citation counts may decrease if, for some reason, a group of citing documents becomes unavailable in the Web.

In order to understand this phenomenon, we may observe the case of the most cited document in the sample (See Table I), which is Lowry's article: "Protein measurement with the Folin phenol reagent". This study suffered a severe drop in citations in the space of a few months. We observed the number of citations of this article at three different points in time: 28th May; 7th August; 21st October, 2014. As of the 28th of May, 2014 (first data sample), it was the most cited document in our sample, with 253,671 citations according to GS. However, on the 21st of October, its citation count had decreased to 192,841 (Table V).

**Table V. Fluctuation of citations received by Lowry's article**

| Date | WoS Citations | GS Citations | Screenshots |
|---|---|---|---|
| 28$^{th}$ May, 2014 | 303,832 | 253,671 | |
| 7$^{th}$ August, 2014 | 304,667 | 191,669 | |
| 21$^{st}$ October, 2014 | 305,202 | 192,841 | |

Within 5 months, Lowry's article lost approximately 60,000 citations. As a consequence, as of October, 2014, it was not the highest cited article in GS, giving way to Laemli's work, which had 223,264 citations. WoScc data seems to be much more stable, showing 303,832 citations in May and 305,202 in October. Conversely, "Diagnostic and statistical manual of mental disorders" (5$^{th}$ position), reported 129,473 citations in May whereas in October the count increased to 185,000 citations, that is, 55,170 more citations in just 5 months.

Presumably, this drastic change in citations took place as a consequence of a major "re-crawling" performed by Google in June 2014. In any case, we believe that this variability may affect specific positions in the ranking of Table I, but not the condition of the documents as highly-cited documents (especially in the top 1%). Of course, this phenomenon is likely to occur on highly cited items, as the number of their citations follows a skewed distribution. The impact of these errors could be large however for non-highly cited items (usual search results).

*The accuracy of duplicate detection / merging versions*

Google Scholar declares that they merge all versions of a same document (not only different editions or reprints published in different years but also translations to other languages), and that all their respective citations are then added (Verstak & Acharya, 2013). However, this task isn't always accomplished successfully. In Figure 9 we can see an example of two different editions (English and Spanish) for the seminal work "Degeneration and regeneration of the nervous system" by Santiago Ramón y Cajal, which haven't been merged. Even for editions in the same language, several variants can be found as well.

**Figure 9. Example of language versions (English and Spanish) of "Degeneration and regeneration of the nervous system by Cajal in Google Scholar**

This simple test suggests that book impact, measured through citations from Google Scholar, would likely be even higher if all versions were successfully merged. This would probably mean that even more books would appear in Table I.

To understand the extent of the issue of citations to a given work which are dispersed among several duplicate records, we carried out a systematic and exhaustive analysis of one book as a case study: "The Mathematical Theory of Communication", by Shannon and Weaver. This work, because of its bibliographic complexity, illustrates the challenges that the correct treatment of highly-cited documents would pose (See supplementary material).[1]

"A mathematical theory of communication" was first published by Shannon as a two-part article in 1948. This work was later expanded and reedited in book form in 1949. This new edition was co-authored by Shannon and Weaver, with a slightly different title: "(The) mathematical theory of communication". Therefore, technically there are two distinct citable items which, ideally, Google Scholar should have been able to tell apart at the moment they were indexed.

In order to learn how GS actually handled this work, we searched it with the query <"mathematical theory of communication"> and selected the result with the greater number of detected versions (830), which we will call the "main record". We downloaded the bibliographic information of all the versions GS found for the main record, which weren't actually 830, but only 763 (discrepancies between hit counts and the actual visible results are a well-known phenomenon in GS).

After this, we refined this query (adding the search command <author:Shannon>) obtaining 229 additional results. Of them, only 164 (71.6%) were actually different versions of the work. The rest were comments and reviews. These 164 records are duplicates that Google Scholar should also have merged with the main record (added to those 763 versions), but didn't.

If we consider the 165 verified records (the main record and the 164 duplicates), the main record held the larger number citations (69,738), whereas the remaining 164 duplicates together accounted for 3,714 new potential citations (not considering possible duplicates or false citations).

This analysis (search, download, and manual check) was carried out in October 2014. A complete description is provided in the supplementary material.[1]

There is a low Pearson's correlation between the number of citations and the number of versions (r= 0.2; n= 64000). This value is similar to that obtained by Jamali and Nabavi (2015), who found a weak positive correlation between the number of versions and the citation counts for full-text articles (r = 0.346; n = 4426). Pitol and De Groote (2014) found low values as well (r= 0.257; n= 982) when describing the GS versions for articles stored in institutional repositories from three US universities.

However, we found that this correlation increases when the Spearman method is used instead (r= 0.48; n= 64000), probably revealing a threshold beyond which it is unusual to find documents with a high number of versions and low citation counts. This result may also indicate that the number of missing citations (from undetected duplicates) will only be significant for highly-cited documents with a high number of versions, which in any case constitute a small portion of the records (they are mainly books). Therefore, there shouldn't be many highly-cited documents that haven't made it to our sample because of Google Scholar's duplicate detection errors.

**Composition of the sample**

*Publication date*

In Table I (highly-cited documents) we can see that the eleventh position is held by a book published outside the timeframe selected in our study (1948). This book, however, appeared in the results of the different queries we performed. Additionally, in Figure 3 we detected an uncommon increment of the presence of books in the results GS displayed for the most recent years. These issues led us to question the information about the publication date that Google Scholar provides for books.

We realized that Google Scholar lumps together all the different editions of the same book, and usually (not always) selects the latest edition as the primary version, taking the date of this version as the publication date of the book. This is the reason behind the fact that the seminal work "A mathematical theory of communication" published by Shannon in 1948 is included in the sample: Google Scholar has selected a reprint published in 2001 as the primary version.

Since Google Scholar only presents 1,000 results for any given query (and we only collected information about the primary versions of the documents), new editions of old books took the place of other publications that had really been published in those years.

The differences between the date of the first edition and the publication date used by Google Scholar for each book is shown in Figure 10 for the top 600 most cited books, where a bias in the last 10 years is evident.

**Figure 10. Number of books according to the year of publication signed by Google Scholar and to the date of the first edition (top 600)**

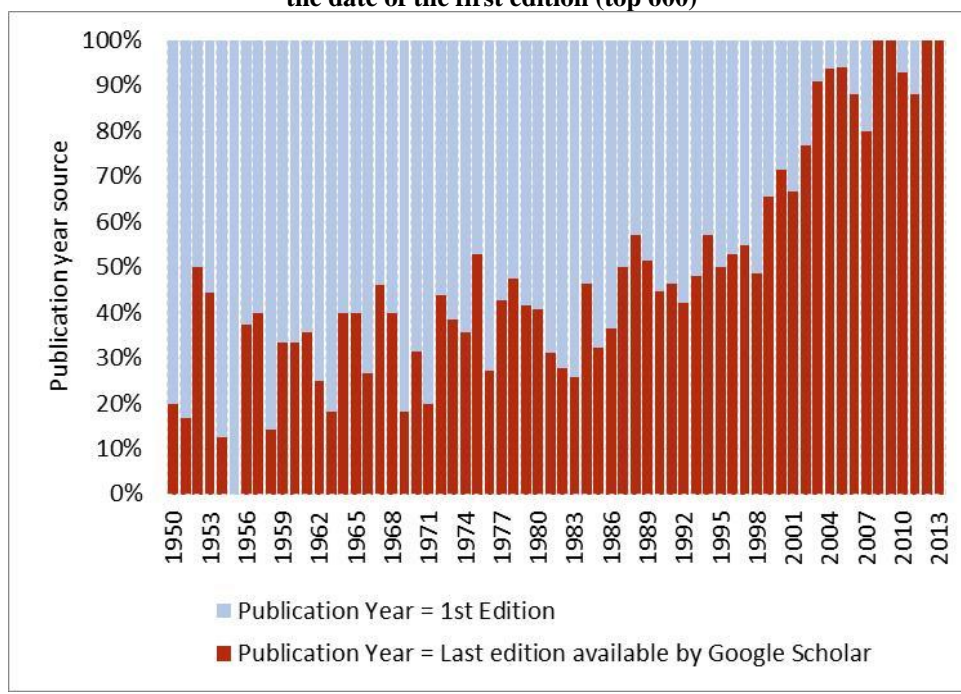

The decision to select the publication date of the most recent edition of a book as the date of publication of the primary version makes a lot of sense from the point of view of a search engine (users will probably want to access the latest edition of a book), but it becomes a problem when the goal is to perform any kind of bibliometric analysis. This issue obviously affects our sample (it is especially noticeable in figure 2 and 3). In any case, it should be noted that this limitation doesn't affect the status of these books as highly-cited documents; only the year of publication is affected, resulting in an overrepresentation of books in the last decade, which are unfairly taking the place of other highly-cited documents that were actually published in those years.

*Language of the documents*

We developed a strategy to determine this information using WoScc data where possible (around 50% of the sample) as well as the title and abstract of the document in all the other cases. This approach, however, may have resulted in an overrepresentation of the English language, since it is usual for a document written in a language other than English to provide its title and abstract in English as well, for the purpose of being indexed in international databases.

Additionally, the sample may contain records that are in fact translations of other documents (which may also be present in our sample). This is the case of journals that are published both in English and in other language or books that are translated into various languages.

For this reason, the English percentage of highly-cited documents should be taken with caution and be considered only as an estimate.

*Highly cited documents*

The selection of only the most cited publications may introduce a bias. So it is possible that these documents, because they are only highly-cited items, would not entirely be representative of all the documents indexed in GS (for example, it is possible that highly-cited papers have more versions, or there are more open full text copies). This would explain their differences with other works. Therefore, the results obtained cannot be extrapolated to the whole Google Scholar database.

*Custom range*

The Google Scholar's custom range option was utilized in order to perform the 64 annual queries. This functionality does not accurately supports Boolean queries and presents some limitations when it comes to retrieve results published on a certain date (Orduna-Malea et al, 2015), which may affect slightly the composition of the sample, especially for those documents without date of publication in the metadata. However, while treating only with highly cited documents the effect of this malfunction diminishes.

**Properties of the sample**

The bibliographic data collected for each document (full-text availability, document type, source provider…) always comes from the version of the document Google Scholar considered as the "primary version" (the one that is displayed in the page of results of a query). This fact constitutes a limitation since one document may be freely accessible through various source providers (for example a journal and a repository) and file formats (for example html and pdf file format). For this reason, all the results obtained, especially those included in the sections 3.4, 3.5 and 3.6 must be interpreted with this limitation in mind. Additionally, it should be reminded that all the queries were performed without activating the academic Library subscriptions feature, which would have introduced a bias in the information about full-text source providers.

*Document type*

The great variety of document types included in Google Scholar, as well as the impossibility of filtering by this variable (Bornmann et al., 2009; Aguillo, 2012) makes document type statistics quite difficult. For this reason, three complementary methods were used in this paper to detect the typology of the 64,000 documents in the sample.

We could only determine the document types of 71% of the entire dataset. A manual inspection would have been required to ascertain the typology of the remaining 29% (18,589 documents). We believe the proportion of books and book chapters would have increased if the entire sample had been successfully categorized, since this is the typology that Google Scholar has more trouble identifying.

*Free Full-text*

Since the existence of a full-text link does not guarantee the disposal of the full-text (some links actually refer to publisher's abstracts), the results (40% of the documents had a free full-text link) might be somewhat overestimated. In any case, these values are consistent with those published by Archambault et al. (2013), who found that over 40% of the articles from their sample were freely accessible; higher than those by Khabsa and Giles (2014) and Björk et al. (2010), who found only a 24% and 20.4% of open access documents respectively; and much lower than Jamali and Nabavi (2015) and Pitol and De Groote (2014), who found 61.1% and 70% respectively.

The different nature of the samples makes it difficult to draw comparisons among these studies. Nonetheless, the sample used in this study (64,000 documents) is the largest ever used to date.

*File format*

The predominance of the pdf and the html file formats matches the results thrown by previous studies. Among others, those by Orduna-Malea et al. (2010), Aguillo et al. (2010), and Jamali and Nabavi (2015).

*Source providers*

The source providers for freely accessible highly-cited documents in Google Scholar are, at least as far as our sample is concerned, institutional (US universities) and subject (Pubmed central and Arxiv) repositories. Despite the fact that some commercial publishers also appear on the top positions of the ranking of source providers, their presence in absolute numbers is small. Of special note is the role of the scientific social network ResearchGate. Its presence, already noted by Jamali and Nabavi (2015), shows that a) ResearchGate contains an already large (and still growing) percentage of highly-cited documents; and b) its capacity to become the primary version of the highly-cited documents in Google Scholar.

These results differ from those obtained by Ortega (2014) who detected a high presence of publishers (constituting the source for 58.4% of all scientific documents in Google Scholar). The reason behind this difference is that Ortega used <site:> queries directly to find the number of documents hosted within the source providers' websites. The different way in which we conducted our queries makes a direct comparison impossible, but it confirms that even though most publishers now allow Google Scholar to crawl their websites, they are not becoming the main destination for users to access the full-text of highly-cited documents.

Regarding the web domains, Aguillo (2012) detected countries which intensely contribute to increase the size of Google Scholar (such as France, Japan, Brazil or China). However, these countries do not appear as the main contributors of highly-cited documents (Germany is the first country in this ranking). The comparison of a general ranking of source providers and the source providers of the highly-cited documents might serve to identify the places where these top contributions actually become freely available to final users on the Web.

Finally, although the existence of commercial agreements with some publishers (information undisclosed by Google Scholar) as well as the development of some Google Scholar's optimization techniques may influence the global coverage, their effect in a sample of most cited documents is estimated to be low. Otherwise, the irregular coverage according to disciplines (not all knowledge areas are equally covered) might disfavour some fields. Notwithstanding, this research is based on what Google Scholar is capable to index. Those contents not indexed due to both technical limitations and specific web policies are excluded.

## 5. CONCLUSIONS

In light of the results obtained, we can conclude that Google Scholar offers an original and different vision of the most influential documents in the academic/scientific environment (measured from the perspective of their citation count). These results are a faithful reflection of the all-encompassing indexing policies that enable Google Scholar to retrieve a larger and more diverse number of citations, since they come from a wider range of document types, different geographical environments, and languages.

Therefore, Google Scholar covers not only seminal research works in the entire spectrum of the scientific fields, but also the greatly influential works that scientists, teachers and professionals who are training to become practitioners use in their respective fields. This phenomenon is particularly true for works that deal with new data collecting and processing techniques.

This is reflected on the high proportion of books among the highly cited documents (62% of the top 1% most cited documents collected), as this document type is essential in the humanities and the social sciences (also as a vehicle for the communication of new results), and in the experimental sciences (as a way to consolidate and disseminate scientific knowledge).

There are still important limitations and errors when working with data extracted from Google Scholar, especially those related to the detection of duplicate documents, and the correct allocation of citations. These issues have all been discussed in-depth in this study. While these mistakes may introduce biases in the ranking of most-cited documents in Google Scholar (the specific position of a document in this list), our empirical data suggest that the influence of these errors on the characterization and description of the sample, which is the main goal of this study, would be minimal.

In conclusion, thanks to the wide and diverse list of sources from which Google Scholar feeds, this search engine covers academic documents in a broader sense, enabling the measurement of impact stemming not only from the scientific side of the academic landscape, but also from the educational side (doctoral dissertations, handbooks) and from the professional side (working papers, technical reports, patents), the last two being areas that haven't been explored as much as the first one.

Other specific findings of this study are summarized below:
- 40% of the highly cited documents in Google Scholar are freely accessible, mostly from educational institutions (mainly universities), and other non-profit organizations.

- Google Scholar has detected more than one version for 83.17% of the documents in our sample.
- The general correlation between the number of versions and the number citations they have received is low (r= 0.2) except for documents with a very high number of versions (more than 100), which also present a high number of citations.
- The average highly-cited document is a journal article (72.3% of the documents for which a document type could be ascertained) or a book (62% of the top 1% most cited documents of the sample), written in English (92.5% of all documents) and available online in PDF format (86.0% of all documents)

## 6. ACKNOWLEDGEMENTS

Research partially funded under the project HAR2011-30383-C02-02, from Dirección General de Investigación y Gestión del Plan Nacional de I+D+I (Ministry of Economy and competitiveness), and the project APOSTD/2013/002, from Conselleria de Educación, Cultura y Deporte (Generalitat Valenciana), in Spain.

## 7. ENDNOTES

1. Supplementary material. Available at https://dx.doi.org/10.6084/m9.figshare.1224314.v1. Accessed 25 March 2016.

## 8. REFERENCES

Aguillo, Isidro F.; Ortega, J.; Fernández, M.; Utrilla, A. (2010). Indicators for a webometric ranking of open access repositories. *Scientometrics*, vol. 82 (3), 477-486. http://dx.doi.org/10.1007/s11192-010-0183-y

Aguillo, Isidro F. (2012). Is Google Scholar useful for bibliometrics? A webometric analysis. *Scientometrics*, vol. 91 (2), 343-351. http://dx.doi.org/10.1007/s11192-011-0582-8

Aksnes, D. W. (2003). Characteristics of highly cited papers. *Research Evaluation*, vol. 12 (3), 159-170. http://dx.doi.org/10.3152/147154403781776645

Aksnes, D. W.; Sivertsen, G. (2004). The effect of highly cited papers on national citation indicators. *Scientometrics*, vol. 59 (2), 213-224. http://dx.doi.org/10.1023/b:scie.0000018529.58334.eb

Archambault, E.; Amyot, D.; Deschamps, P.; Nicol, A.; Rebout, L.; Roberge, G. (2013). Proportion of open access peer-reviewed papers at the European and world levels—2004–2011. Science-Metrix. Report. Science Matrix Inc. Disponible en: http://www.science-metrix.com/pdf/SM_EC_OA_Availability_2004-2011.pdf

Bar-Ilan, J. (2010). Citations to the "Introduction to informetrics" indexed by WOS, Scopus and Google Scholar. *Scientometrics*, vol. 82(3), 495-506. http://dx.doi.org/10.1007/s11192-010-0185-9

Beel, J.; Gipp, B.; Wilde, E. (2010). Academic Search Engine Optimization (ASEO): Optimizing Scholarly Literature for Google Scholar and Co. *Journal of Scholarly Publishing*, vol. 41 (2), 176-190. http://dx.doi.org/10.1353/scp.0.0082

Björk, B. C.; Welling, P.; Laakso, M.; Majlender, P.; Hedlund, T.; Gudnason, G. (2010). Open Access to the scientific journal literature: Situation 2009. *PLoS ONE*, vol. 5(6), e11273. http://dx.doi.org/10.1371/journal.pone.0011273

Bornmann, L. (2010). Towards an ideal method of measuring research performance: Some comments to the Opthof and Leydesdorff (2010) paper. *Journal of Informetrics*, vol. 4 (3), 441–443. http://dx.doi.org/10.1016/j.joi.2010.04.004

Bornmann, L.; Mutz, R. (2011). Further steps towards an ideal method of measuring citation performance: the avoidance of citation (ratio) averages in field-normalization. *Journal of Informetrics*, vol. 5 (1), 228-230. http://dx.doi.org/10.1016/j.joi.2010.10.009

Bornmann, L.; Marx, W.; Schier, H.; Rahm, E.; Thor, A.; Daniel, H. D. (2009). Convergent validity of bibliometric Google Scholar data in the field of chemistry—Citation counts for papers that were accepted by Angewandte Chemie International Edition or rejected but published elsewhere, using Google Scholar, Science Citation Index, Scopus, and Chemical Abstracts. *Journal of Informetrics*, vol. 3 (1), 27-35. http://dx.doi.org/10.1016/j.joi.2008.11.001

Bornmann, L.; Moya-Anegón, F.; Leydesdorff, L. (2011). The new excellence indicator in the World Report of the SCImago Institutions Rankings 2011. *Journal of Informetrics*, vol. 6(2), 333-335. http://dx.doi.org/10.1016/j.joi.2011.11.006

Garfield, E. (1977). Introducing Citation Classics: the human side of scientific papers. *Current contents*, vol. 3 (1), 1-2.

Garfield, E. (1979). Is citation analysis a legitimate evaluation tool?. *Scientometrics*, vol. 1(4), 359-375.

Garfield, E. (2005). The Agony and the Ecstasy—The History and Meaning of the Journal Impact Factor. *International Congress on Peer Review and Biomedical Publication*. Chicago, 16 September. Disponible en: http://www.garfield.library.upenn.edu/papers/jifchicago2005.pdf

Glänzel, W.; Czerwon, H. J. (1992). What are highly cited publications? A method applied to German scientific papers, 1980–1989. *Research Evaluation*, vol. 2 (3), 135-141. http://dx.doi.org/10.1093/rev/2.3.135

Glänzel, W.; Schubert, A. (1992). Some facts and figures on highly cited papers in the sciences, 1981–1985. *Scientometrics*, vol. 25 (3), 373-380. http://dx.doi.org/10.1007/bf02016926

Glänzel, W.; Rinia, E. J.; Brocken, M. G. (1995). A bibliometric study of highly cited European physics papers in the 80s. *Research Evaluation*, vol. 5 (2), 113-122. http://dx.doi.org/10.1093/rev/5.2.113

Harzing, A. W. (2013). A preliminary test of Google Scholar as a source for citation data: a longitudinal study of Nobel prize winners. *Scientometrics*, vol. 94 (3), 1057-1075. http://dx.doi.org/10.1007/s11192-012-0777-7

Harzing, A. W. (2014). A longitudinal study of Google Scholar coverage between 2012 and 2013. *Scientometrics*, vol. 98 (1), 565-575. http://dx.doi.org/10.1007/s11192-013-0975-y

Harzing, A.W.; Van der Wal, R. (2008). Google Scholar as a new source for citation analysis. *Ethics in Science and Environmental Politics*, vol. 8 (1), 61-73. http://dx.doi.org/10.3354/esep00076

Jacso, P. (2005). Google Scholar: the pros and the cons. *Online information review*, vol. 29 (2), 208-214. http://dx.doi.org/10.1108/14684520510598066

Jacso, P. (2006). Deflated, inflated, and phantom citation counts. *Online Information Review*, vol. 30 (3), 297-309. http://dx.doi.org/10.1108/14684520610675816

Jacsó, P. (2008a). The pros and cons of computing the h-index using Scopus. *Online Information Review*, vol. 32 (4), 524-535. http://dx.doi.org/10.1108/14684520810897403

Jacso, P. (2008b). The pros and cons of computing the h-index using Google Scholar. *Online Information Review*, vol. 32 (3), 437-452. http://dx.doi.org/10.1108/14684520810889718

Jacso, P. (2012). Using Google Scholar for journal impact factors and the h-index in nationwide publishing assessments in academia – siren songs and air-raid sirens. *Online Information Review*, vol. 36 (3), 462-478. http://dx.doi.org/10.1108/14684521211241503

Jamali, H. R. ; Nabavi, M. (2015). Open access and sources of full-text articles in Google Scholar in different subject fields. *Scientometrics*, vol. 105 (3), 1635-1651. http://dx.doi.org/10.1007/s11192-015-1642-2

Khabsa, M.; Giles, C. L. (2014). The number of scholarly documents on the public web. *PLoS One*, vol. 9(5), e93949. http://dx.doi.org/10.1371/journal.pone.0093949

Kousha, K.; Thelwall, M. (2008). Sources of Google Scholar citations outside the Science Citation Index: A comparison between four science disciplines. *Scientometrics*, vol. 74 (2), 273–294. http://dx.doi.org/10.1007/s11192-008-0217-x

Kousha, K.; Thelwall, M.; Rezaie, S. (2011). Assessing the citation impact of books: The role of Google Books, Google Scholar, and Scopus. *Journal of the American Society for Information Science*, vol. 62 (11), 2147-2164. http://dx.doi.org/10.1002/asi.21608

Kresge, N.; Simoni, R. D.; Hill, R. L. (2005). The most highly cited paper in publishing history: Protein determination by Oliver H. Lowry. *Journal of Biological Chemistry*, vol. 280 (28), e25. Disponible en: http://www.jbc.org/content/280/28/e25

Levitt, J. M.; Thelwall, M. (2009). The most highly cited Library and Information Science articles: Interdisciplinarity, first authors and citation patterns. *Scientometrics*, vol. 78 (1), 45-67. http://dx.doi.org/10.1007/s11192-007-1927-1

Maltrás Barba, B. (2003). *Los indicadores bibliométricos: fundamentos y aplicación al análisis de la ciencia*. Gijón: Trea.

Martín-Martín, A.; Ayllón, J. M.; Delgado López-Cózar, E.; Orduna-Malea, E. (2015). Nature's top 100 Re-revisited. *Journal of the Association for Information Science & Technology*, vol. 66 (12), 2714-2714. http://dx.doi.org/10.1002/asi.23570

Meho, L.; Yang, K. (2007). Impact of data sources on citation counts and rankings of LIS faculty: Web of Science versus Scopus and Google Scholar. *Journal of the American Society for Information Science and Technology*, vol. 58 (13), 2105–2125. http://dx.doi.org/10.1002/asi.20677

Narin, F. (1987). Bibliometric techniques in the evaluation of research programs. *Science and Public Policy*, vol. 14(2), 99-106.

Narin, F.; Frame, J. D.; Carpenter, M. P. (1983). Highly cited Soviet papers: An exploratory investigation. *Social Studies of Science*, vol. 13 (2), 307-319. http://dx.doi.org/10.1177/030631283013002006

Oppenheim, C.; Renn, S. P. (1978). Highly cited old papers and the reasons why they continue to be cited. *Journal of the American Society for Information Science*, vol. 29 (5), 225-231. http://dx.doi.org/10.1002/asi.4630290504

Orduna-Malea, E.; Delgado López-Cózar, E. (2014). Google Scholar Metrics evolution: an analysis according to languages. *Scientometrics*, vol. 98 (3), 2353–2367. http://dx.doi.org/10.1007/s11192-013-1164-8

Orduna-Malea, E.; Ayllón, J. M.; Martín-Martín, A.; Delgado López-Cózar, E. (2015). Methods for estimating the size of Google Scholar. *Scientometrics*, vol. 104 (3), 931-949. http://dx.doi.org/10.1007/s11192-015-1614-6

Orduna-Malea, E.; Serrano-Cobos, J.; Ontalba-Ruipérez, J. A.; Lloret-Romero, N. (2010). Presencia y visibilidad web de las universidades públicas españolas. *Revista española de documentación científica*, vol. 33 (2), 246-278. http://dx.doi.org/10.3989/redc.2010.2.740

Ortega, Jose L. (2014). *Academic Search Engines: A Quantitative Outlook*. London: Elsevier.

Persson, O. (2010). Are highly cited papers more international?. *Scientometrics*, vol. 83 (2), 397-401. http://dx.doi.org/10.1007/s11192-009-0007-0

Pitol, S. P. ; De Groote, S. L. (2014). Google Scholar versions: Do more versions of an article mean greater impact? *Library Hi Tech*, vol. 32 (4), 594–611. http://dx.doi.org/10.1108/lht-05-2014-0039

Plomp, R. (1990). The significance of the number of highly cited papers as an indicator of scientific prolificacy. *Scientometrics*, vol. 19 (3), 185-197. http://dx.doi.org/10.1007/bf02095346

Smith, D. R. (2009). Highly cited articles in environmental and occupational health, 1919–1960. *Archives of environmental & occupational health*, vol. 64 (1), 32-42. http://dx.doi.org/10.1080/19338240903286743

Tijssen, R. J.; Visser, M. S.; Van Leeuwen, T. N. (2002). Benchmarking international scientific excellence: are highly cited research papers an appropriate frame of reference? *Scientometrics*, vol. 54 (3), 381-397.

Van Noorden, R.; Maher, B.; Nuzzo, R. (2014). The top hundred papers. *Nature*, vol. 514 (7524), 550-553. http://dx.doi.org/10.1038/514550a

Van Raan, A. F.; Hartmann, D. (1987). The comparative impact of scientific publications and journals: Methods of measurement and graphical display. *Scientometrics*, vol. 11(5-6), 325-331. http://dx.doi.org/10.1007/bf02279352

Verstak, A.; Acharya, A. (2013). *Identifying multiple versions of documents*. U.S. Patent No. 8,589,784. Washington, DC: U.S. Patent and Trademark Office.

Winter, J.C.F.; Zadpoor, A.; Dodou, D. (2014). The expansion of Google Scholar versus Web of Science: a longitudinal study. *Scientometrics*, vol. 98 (2), 1547–1565. http://dx.doi.org/10.1007/s11192-013-1089-2

Yang, K.; Meho, L. (2006). Citation Analysis: A Comparison of Google Scholar, Scopus, and Web of Science. *Proceedings of the American Society for Information Science and Technology*, vol. 43 (1), 1–15. http://dx.doi.org/10.1002/meet.14504301185